\documentclass[sigconf]{acmart}

\AtBeginDocument{%
  \providecommand\BibTeX{{%
    \normalfont B\kern-0.5em{\scshape i\kern-0.25em b}\kern-0.8em\TeX}}}

\copyrightyear{2020}
\acmYear{2020}
\setcopyright{acmcopyright}\acmConference[MM '20]{Proceedings of the 28th ACM International Conference on Multimedia}{October 12--16, 2020}{Seattle, WA, USA}
\acmBooktitle{Proceedings of the 28th ACM International Conference on Multimedia (MM '20), October 12--16, 2020, Seattle, WA, USA}
\acmPrice{15.00}
\acmDOI{10.1145/3394171.3413762}
\acmISBN{978-1-4503-7988-5/20/10}

\newcommand{\tabincell}[2]{\begin{tabular}{@{}#1@{}}#2\end{tabular}}
\usepackage{multirow}
\usepackage{graphicx}
\usepackage{subfigure}
\usepackage{stfloats}
\usepackage{balance}

\begin{document}
\fancyhead{}
\title{Towards Modality Transferable Visual Information Representation with Optimal Model Compression}


\author{Rongqun Lin}
\affiliation{%
\institution{Department of Computer Science,\\ City University of Hong Kong}
}
\email{rqlin3-c@my.cityu.edu.hk}

\author{Linwei Zhu}
\affiliation{%
\institution{Shenzhen Institutes of Advanced Technology,\\ Chinese Academy of Sciences}
}
\email{lw.zhu@siat.ac.cn}

\author{Shiqi Wang}
\authornote{Shiqi Wang and Sam Kwong are the corresponding authors.}
\affiliation{%
  \institution{Department of Computer Science,\\ City University of Hong Kong}
}
\email{shiqwang@cityu.edu.hk}

\author{Sam Kwong}
\authornotemark[1]
\affiliation{%
  \institution{Department of Computer Science,\\ City University of Hong Kong}
}
\email{cssamk@cityu.edu.hk}

\renewcommand{\shortauthors}{xxx, et al.}

\begin{abstract}
Compactly representing the visual signals is of fundamental importance in various image/video-centered applications. 
Although numerous approaches were developed for improving the image and video coding performance by removing the redundancies within visual signals, much less work has been dedicated to the transformation of the visual signals to another well-established modality for better representation capability. 
In this paper, we propose a new scheme for visual signal representation that leverages the philosophy of transferable modality. In particular, the deep learning model, which characterizes and absorbs the statistics of the input scene with online training, could be efficiently represented in the sense of rate-utility optimization to serve as the enhancement layer in the bitstream. As such, the overall performance can be further guaranteed by optimizing the new modality incorporated. The proposed framework is implemented on the state-of-the-art video coding standard (i.e., versatile video coding), and significantly better representation capability has been observed based on extensive evaluations.
\end{abstract}



\maketitle

\section{Introduction}
Recently, we have witnessed exponential growth of image/video services, coinciding with the accelerated proliferation of acquisition and display devices. The gigantic scale of visual data motivates the research of  compact signal representation, which is the long-standing problem and  indispensable in numerous applications. Traditional methods aim to remove the redundancies within the visual signals, such as spatial, temporal, statistical and perceptual redundancies. Based on the philosophy of redundancy removal, a series of video coding standards have been developed, including
H.264/AVC \cite{wiegand2003overview}, H.265/HEVC \cite{sullivan2012overview}, VP9\cite{mukherjee2013technical}, AV1\cite{chen2018overview}, H.266/VVC \cite{bross2018working} and AVS~\cite{avs2Ma}.   
 
With the surge of deep learning, numerous efforts have been devoted to improving the compact signal representation capability with deep neural networks, including incorporating the neural network into the hybrid video coding  framework~\cite{li2018fully,li2018hybrid,pfaff2018intra,li2018convolutional,zhang2017efficient,hu2018enhanced,huo2018convolutional,yan2018convolutional,zhao2018enhanced,zhao2019enhanced,song2017neural, meng2018new, park2016cnn, zhang2018residual,ding2019switchable,jia2017spatial,jia2019content}  and end-to-end compression~\cite{toderici2015variable,balle2015density,balle2016end,balle2016end2,rippel2017real,agustsson2019generative,tschannen2018deep,santurkar2018generative}. In the first category, intra prediction, inter prediction, loop filtering and entropy coding modules have been significantly enhanced with the deep neural networks at both video encoder and decoder sides. In the second category, the visual information is compactly represented with the latent code in the manner of end-to-end training. Though promising performance has been achieved, the systematical study regarding the redundancy removal by transferring from visual information to deep learning model  which is recognized as one important modality of knowledge \cite{chen2018data} on data statistics has been largely ignored. 

The deep neural networks have been regarded as the important modality of knowledge in Knowledge Centric Networking (KCN)~\cite{wu2017vision}, and the network communication has been widely studied in the literature~\cite{chen2019toward}. In video delivery~\cite{yeo2017will,yeo2018neural}, the network has been learned and coded in an online manner, which significantly improves the performance of video streaming. In this paper, based on the extensive studies on quality enhancement with neural networks, we make a further attempt by optimally transferring the visual signals to another well-established modality deep neural network for better signal representation capability. We aim to explore the possibility of efficient representation of the visual information with deep learning model in the sense of rate-utility optimization, such that the model information could serve as an enhancement layer in the representation bitstream. As such, instead of internally removing the redundancies of the visual content, the visual information is further represented with the assistant of the knowledge in deep learning models. The contributions of this paper are summarized as follows.
\begin{itemize}
\item We propose to leverage the representation capability of deep neural networks for further visual redundancy removal on top of the state-of-the-art compression framework. The proposed scheme is designed based on the philosophy of online modality transfer with model compression and optimal model selection. 
\item We propose to efficiently compress the deep learning model with an effort to optimize the transferable modality. Instead of quantizing the weight after online model training, weight quantization has been incorporated  by the scale transform and affine transform during the online training, such that the model representation is optimized in the training phase, leading to reproducible performance in the testing phase. 
\item We propose to optimize the model representation with rate-utility optimization. In particular, instead of only ensuring the optimal signal representation capability, the model rate is also considered in the optimization process. As such, the overall performance can be ensured by optimal model selection and representation.  
\end{itemize}

\section{Related works}

\subsection{Image/video compression}

Traditional compact visual information representation relies on image/video compression, and recently
numerous image/video coding standards have been developed based on the hybrid coding framework, such as JPEG~\cite{wallace1990overview}, 
H.264/AVC ~\cite{wiegand2003overview}, H.265/HEVC ~\cite{sullivan2012overview}, VP9~\cite{mukherjee2013technical}, AV1~\cite{chen2018overview}, AVS~\cite{avs2Ma} and H.266/VVC~\cite{bross2018working}. 
In these standards, the spatial, temporal, and statistical redundancies have been fully exploited to improve the coding performance. 

Regarding spatial redundancy, the state-of-the-art intra coding extends the number of angular prediction modes to 67~\cite{Modes67} for better capturing the arbitrary texture directions. Moreover, Multiple Reference Line (MRL)~\cite{MultiRef} intra prediction is adopted where more informative reference lines are involved in the conventional intra prediction procedure, leading to further removal of the spatial redundancy. 
To remove temporal redundancy, affine motion compensated prediction~\cite{vvc4} has been introduced to improve the precision of irregular motions, such as zoom in/out and rotation.

For the removal of the statistical redundancy, Context-Adaptive Binary Arithmetic Coding (CABAC)~\cite{HEVC_CABAC} is an efficient entropy coding method, which has been used in H.264/AVC, H.265/HEVC, AVS, VVC etc. 
CABAC combines the adaptive binary arithmetic coding with the context modeling, which brings sufficient adaptation and redundancy reduction in a lossless way. 
Additionally, Adaptive Loop Filter (ALF)~\cite{vvc4} is placed at the last stage of the codec, which is a Wiener-filter targeting at minimizing the mean squared error between the original and reconstructed frames. 
The entire coding process is optimized with rate-distortion optimization \cite{sullivan1998rate} to ensure the optimal coding performance.

\subsection{Neural network compression}

Recently, numerous efforts have been devoted to neural network compression \cite{cheng2017survey}, aiming to lessen the massive cost of deep neural network in terms of both storage and computation without significant degradation on the performance. Typically, these  approaches can be divided into eight categories: 1) parameters pruning and filter selection, 2) quantization, 3) matrix factorization, 4) transferred convolutional filters, 5) knowledge transfer and distillation, 6) network redesign, 7) transparent compression and 8) entropy constrains. 

In particular, parameter pruning \cite{han2015deep,guo2016dynamic,dong2017learning} aims to prune the unnecessary or weak response neurons, and filter selection approaches \cite{he2017channel,wang2017beyond,kim2015compression,ye2018rethinking} attempt to abolish unimportant channels. Weight quantization \cite{gong2014compressing,courbariaux2015binaryconnect,hubara2016binarized,leng2018extremely,lin2016fixed,park2017weighted,rastegari2016xnor,wu2016quantized,zhou2017incremental,louizos2017bayesian,reagen2017weightless} quantizes the weights of neural network into binary, ternary values or their powers with little degradation on accuracy.  Considering the whole neuron weight as a matrix, matrix factorization \cite{yu2017compressing,masana2017domain,lin2016towards} can further reconstruct the weight matrix with the low rank methods. The design philosophy behind transferred convolutional filters based methods~\cite{shang2016understanding,zhai2016doubly} adopts special structural convolutional filters to shrink the parameters. Regarding the knowledge distilling~\cite{hinton2015distilling,ba2014deep,luo2016face}, a small scale network can be learned under the guidance of a large scale teacher network.  Furthermore, many new network architectures have been proposed by redesigning the network structure,  such as BinaryNet~\cite{courbariaux2016binarized}, XNORNet~\cite{rastegari2016xnor}, SqueezeNet~\cite{iandola2016squeezenet} and MobileNet~\cite{howard2017mobilenets}. Transparent compression method \cite{laude2018neural}  adopts the transform coding strategies without modifying the network structure. Methods based on entropy constrains on parameters \cite{wiedemann2019entropy,oktay2019scalable} utilize entropy penalized policy to produce compact representation of model.

\begin{figure*}[t]
  \centering
  \includegraphics[width=0.65\linewidth]{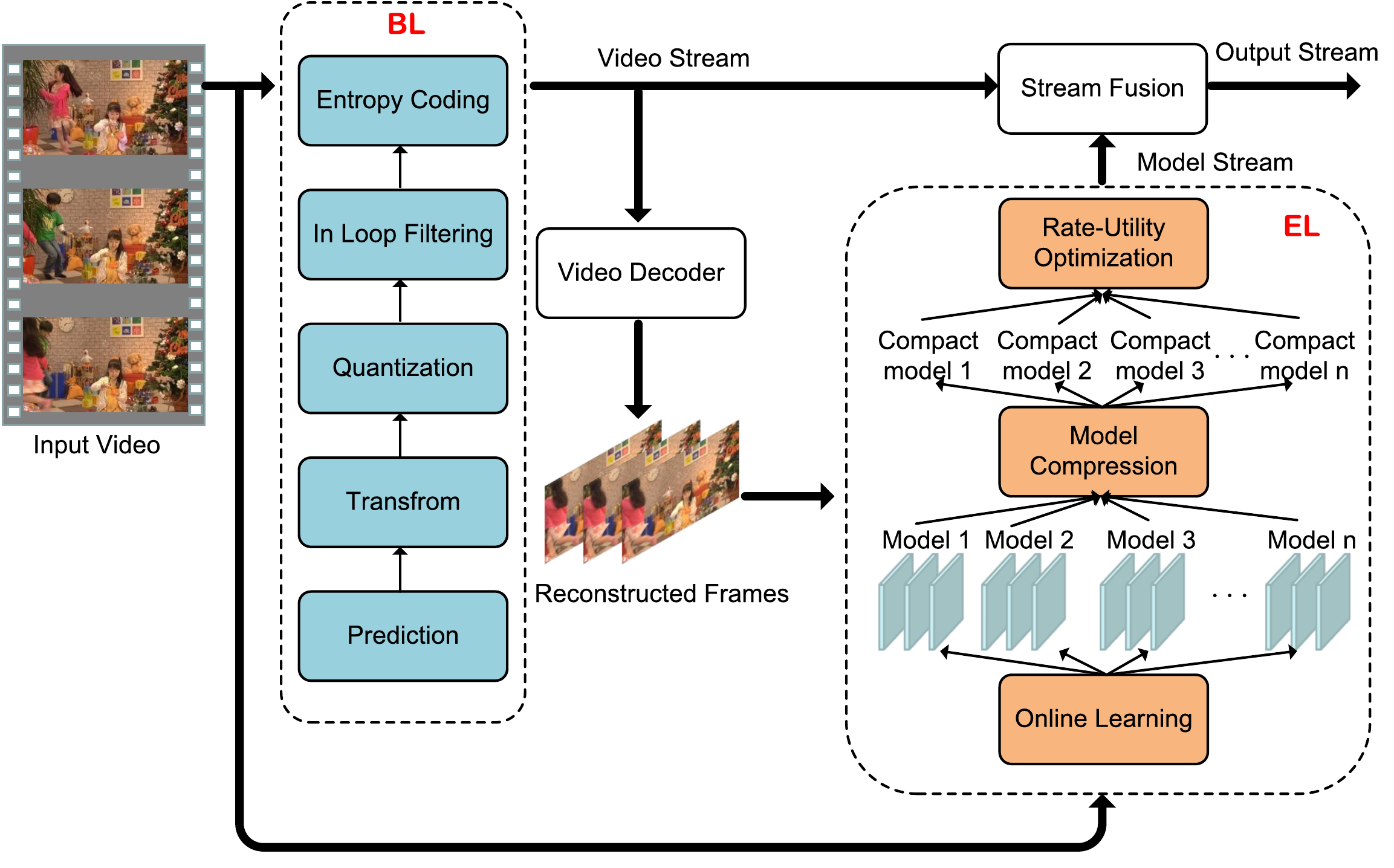}
  \caption{Illustration of the framework of Modality Transferable Visual Information Representation.}
  \Description{The  framework of Modality Transferable Visual Information Representation.}
  \label{fig:framework}
\end{figure*}

\subsection{Deep learning based image/video coding}
With the surge of deep learning in many applications, numerous deep learning based image/video compression approaches have been proposed to achieve more compact representation of visual signals.
The majority of deep learning based video coding approaches fall into two categories: end-to-end compression and the substitution of the modules with deep neural networks in the hybrid coding framework. 

End-to-end compression technologies explore the representation capacities of deep neural networks which could be end-to-end trained. As such, a better trade-off between the bitrate and the distortion can be achieved. The pioneering work \cite{toderici2015variable} utilizes Recurrent Neural Network (RNN) to  reconstruct the image in an end-to-end manner. The Convolutional Neural Network (CNN) based end-to-end image compression methods \cite{balle2015density,balle2016end,balle2016end2} realize effective compression through Generalized Divisive Normalization (GDN) nonlinearity embedded analysis and synthesis transforms.  The pioneering approach utilizing the adversarial
loss function for image compression was proposed by Rippel \textit{et al.} \cite{rippel2017real}. 
Subsequently, the Generative Adversarial Networks (GANs) have been applied to pursue realistic reconstruction quality of images with very low bitrates~\cite{santurkar2018generative,agustsson2019generative,tschannen2018deep}.

The deep neural networks can also be embedded into main modules in the hybrid coding framework to improve the coding performance, 
i.e., intra prediction, inter-prediction, entropy coding and loop filtering. Intra prediction techniques using deep neural networks~\cite{li2018fully,li2018hybrid,pfaff2018intra,li2018convolutional,zhang2017efficient,hu2018enhanced} focus on the improvement of intra prediction efficiency by creating more powerful intra prediction modes.   Deep neural network based inter prediction methods have improved the prediction efficiency \cite{huo2018convolutional,yan2018convolutional,zhao2018enhanced,zhao2019enhanced} by generating a more convincing prediction. Song
\textit{et al.}~\cite{song2017neural} proposed a deep neural network based entropy coding method to directly predict the probability distribution of intra modes instead of relying on the handcrafted context models, such that the statistical redundancy can be further removed, leading to higher coding efficiency. DNNs based loop filtering methods have been widely studied  \cite{ meng2018new, park2016cnn, zhang2018residual,ding2019switchable,jia2017spatial,jia2019content}, which learn the mapping between  the original patches and the degraded patches to eliminate the  inevitable distortion introduced by the block-based hybrid video compression framework. 

In addition, there are a series of approaches incorporating an adaptively learned CNN model in in-loop filtering during the standardlization of VVC~ \cite{JVET-N0110,JVET-N0480}. 
Instead of online learning an adaptive model only, our method aims to explore the capability of modality transferable visual information representation with the design philosophy of rate-utility optimization. 
Moreover, with the proposed rate-utility optimization, the modality transfer capability can be better exploited by incorporating the proposed method with any compact representation frameworks. In view of this, we incorporate the proposed scheme with the state-of-the-art VVC codec as an enhancement layer, and superior coding performance has been achieved. It is noteworthy that the enhancement layer in our method is CNN based and applied on the degraded frames of base layer, such that the proposed scheme is compared with these CNN based video quality enhancement methods on VVC.

\section{Modality Transferable Visual Information Representation}
\subsection{Framework }
As illustrated in Fig. \ref{fig:framework}, the proposed Modality Transferable Visual Information Representation (MTVIR) framework is composed of Base Layer (BL) and Enhancement Layer (EL).
Specifically, the BL aligns with the traditional video codec, including several modules (prediction, transform, quantization, in loop filtering, and entropy coding) which are used to produce compact representation of visual signals based on the removal of intrinsic redundancies. The signal divergences between the original and distorted videos inevitably introduced in the BL are compensated with the modality of deep neural network which is specifically learned by the adaptive transferring of signal level distortion. 

As such, the EL is introduced, which is composed of three sequential modules including online learning, model compression and rate-utility optimization,
and the compact representation of deep neural network is combined with the BL to form the final output stream. 
The framework is able to shift the signal representation to deep learning model representation which is essentially a compact model with enhanced generative capability. Moreover, increased degree of scalability and flexibility is also supported based on the scalable architecture, as the bitstream composed of BL and EL can be adaptively shaped according to the network and storage constrains, and at the decoder side the individual decoding of the BL already supports the reconstruction of the fine texture.

\begin{figure*}[t]
  \centering
  \includegraphics[width=0.8\linewidth]{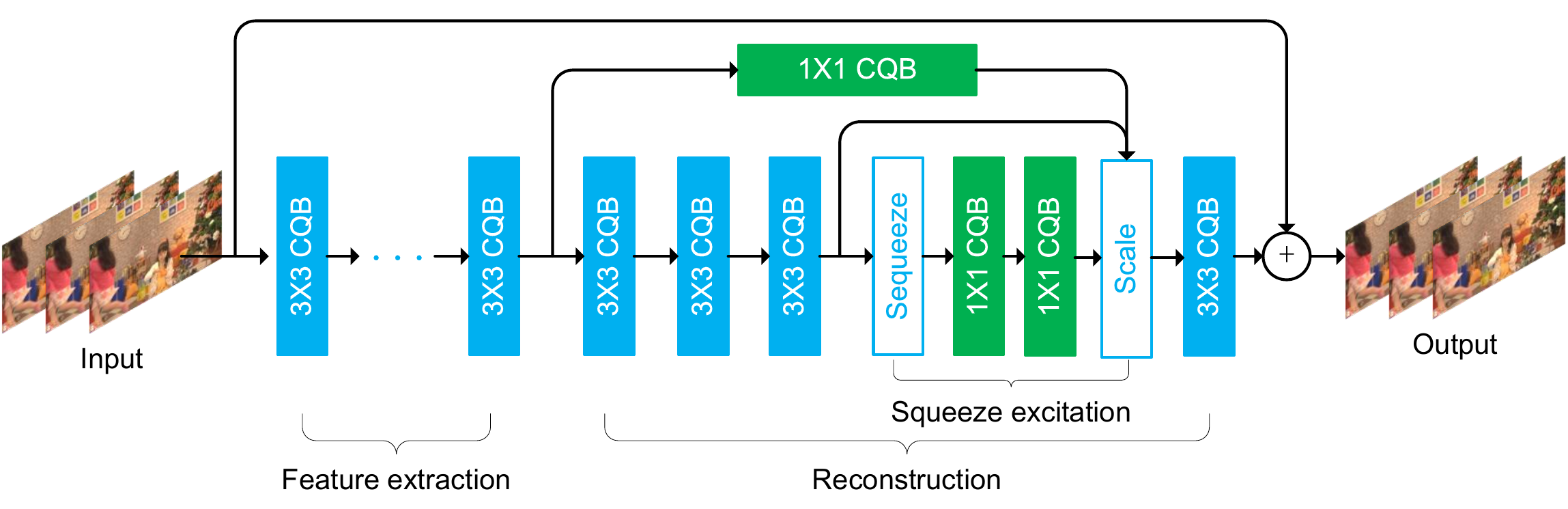}
  \caption{Diagram of the network structure.}
  \Description{}
  \label{fig:network_structure}
\end{figure*}

\begin{figure}[t]
  \centering
  \includegraphics[width=0.8\linewidth]{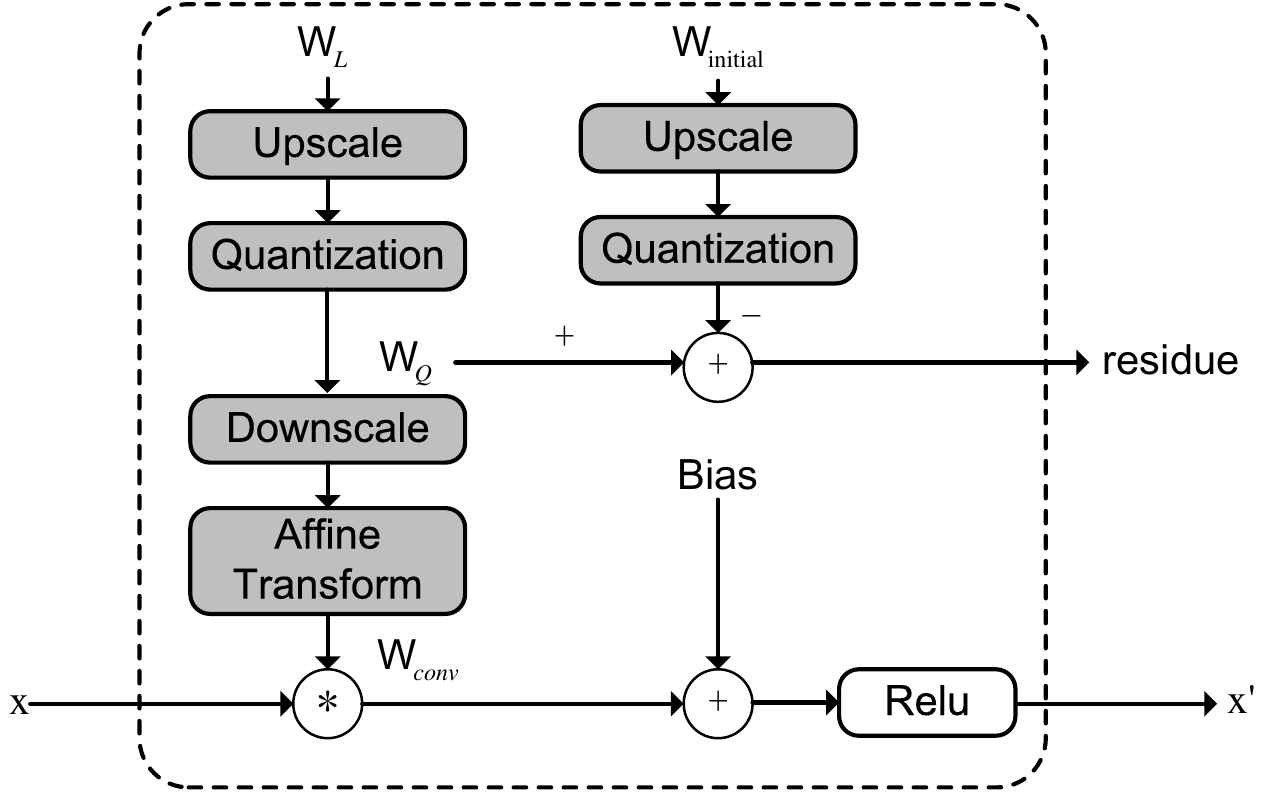}
  \caption{Illustration of the Conv Quantization Block.}
  \Description{}
  \label{fig:conv_quantization_block}
\end{figure}

\subsection{Online learning}
\label{subsection:onlineLearning}
Online learning in EL serves as the transferable engine to characterize and absorb the statistical divergences between the pristine signals and the distorted version, in an effort to transform the visual signals into a well established and highly compact model. 
To this end, a neural network is utilized to learn such a mapping. 
Towards a compact representation, the quantization is performed on the parameters of the neural network with scale transform and affine transform for achieving more compact representation.

In this work, the neural network model is redesigned based upon Squeeze-and-Excitation Filtering CNN (SEFCNN) \cite{ding2019switchable}, as shown in Fig. \ref{fig:network_structure}. In particular, we incorporate the mechanism of compact neural network representation in this model by leveraging the scale transform, quantization operation and affine transform which are integrated as the Conv Quantization Block (CQB),  as shown in Fig. \ref{fig:conv_quantization_block}. In particular, \begin{math} x\end{math} and \begin{math} x'\end{math} represent the input signal and output signal of CQB, respectively.
As such, instead of the convolutional weights and biases learned in SEFCNN, the to-be-learned parameters include the reparameterizations of convolutional weights which are subjected to be quantized, the biases as well as the weights of affine transform. 
More specifically, the scale transform, quantization and affine transform can be formulated as follows,
\begin{equation}
   W_{Q}= round\left ( W_{L}\times S_{c} \right ),\\
\end{equation}
and
\begin{equation}
  W_{conv} =f_{\varphi}\left(W_{Q}\times\frac{1}{S_{c}}\right),
\end{equation}
where \begin{math} W_{Q}\end{math} indicates the quantized weights,  \begin{math} W_{L}\end{math} indicates the weights that should be learned,  \begin{math}  S_{c}\end{math} indicates the scale factor, \begin{math}  W_{conv}\end{math} represents the convolutional weights in network, and $f_{\varphi}(\cdot)$ represents the affine transform.
As can be seen, the quantized weights are obtained by rounding the scaled to-be-learned weights. Then the quantized weights are down-scaled and fed to affine transform to generate the convolutional weights.   The residue between quantized model and  initial quantized model will be compressed by arithmetic coding, which will be discussed in subsection \ref{subsection:model_compression}.

To reduce training computational complexity and enable the residual transmission, the network is trained by fine-tuning through an initial model in online learning. 
More specifically, an initial model with our architecture is obtained by using a large scale dataset and then the online models are learned specifically to fit the statistics of the video signals. 
Since the quantization operation in the network is indifferentiable, the ``straight-through'' gradient estimator proposed by Bengio \textit{et al.}~\cite{bengio2013estimating} is adopted, which performs forward rounding and  backpropagates the gradient directly through the quantization operation to make the network trainable.
In our method, Mean Squared Error (MSE) is adopted as the loss function to pursue the minimization on the difference between the output of the network and the ground truth. 
The to-be-learned weights are updated by minimizing the MSE loss, and subsequently the quantized weights and convolutional weights can be obtained. 
Our network is trained with limited frames, such that the time consumption 
is controllable to a certain extent.

\begin{table}[t]
  \caption{Parameter number comparisons between weights and biases.}
  \label{tab:Parameter_number_comparison}
  \begin{tabular}{cccc}
      \toprule
    Index & Weights &Biases& Total \\
     \midrule
    Parameter number  & 40224& 337 & 40561\\
    Proportion  & 99.17\% & 0.83\%  & 100\% \\
    \bottomrule
\end{tabular}
\end{table}

\begin{table}
  \caption{Parameter setting of biases.}
  \label{tab:paremeter_setting_bias}
  \begin{tabular}{ccc}
      \toprule
    Index &\begin{math} bias^{1},bias^{2},......,bias^{21}\end{math}&\begin{math} bias^{22}\end{math}\\
    \midrule
    Parameter number  & 1$\times$16$\times$21&1\\
   \bottomrule
\end{tabular}
\end{table}

\begin{table*}[h]
  \caption{Parameter setting of weights.}
  \label{tab:paremeter_setting_Weights}
  \begin{tabular}{ccccc}
      \toprule
    Index&\begin{math} W_{conv}^{1}\end{math}&\begin{math}W_{conv}^{2},......, W_{conv}^{18}\end{math}&\begin{math} W_{conv}^{19},W_{conv}^{20},W_{conv}^{21} \end{math}&\begin{math}W_{conv}^{22}\end{math}\\
    \midrule
    Receptive field& 3$\times$3& 3$\times$3& 1$\times$1& 3$\times$3\\
    Feature map number & 16& 16& 16&16\\
    Parameter number&3$\times$3$\times$1$\times$16  & 3$\times$3$\times$16$\times$16$\times$17&1$\times$1$\times$16$\times$16$\times$3&3$\times$3$\times$16$\times$1\\
  \bottomrule
\end{tabular}
\end{table*}

\subsection{Model compression} 
\label{subsection:model_compression}
Model compression aims to compactly represent the learned model by exploiting both intra model redundancy and inter model redundancy.
To remove the redundancy within a model, the quantization of the to-be-learned weights is performed, as detailed in subsection \ref{subsection:onlineLearning}, in an effort to reduce the model size for representation. 
To remove the redundancy across models, the residue between current learned quantized model and the reference quantized model which is universally initialized is encoded to further shrink the model stream.

\begin{figure}[t]
\centering
{\includegraphics[width = 0.35\textwidth]{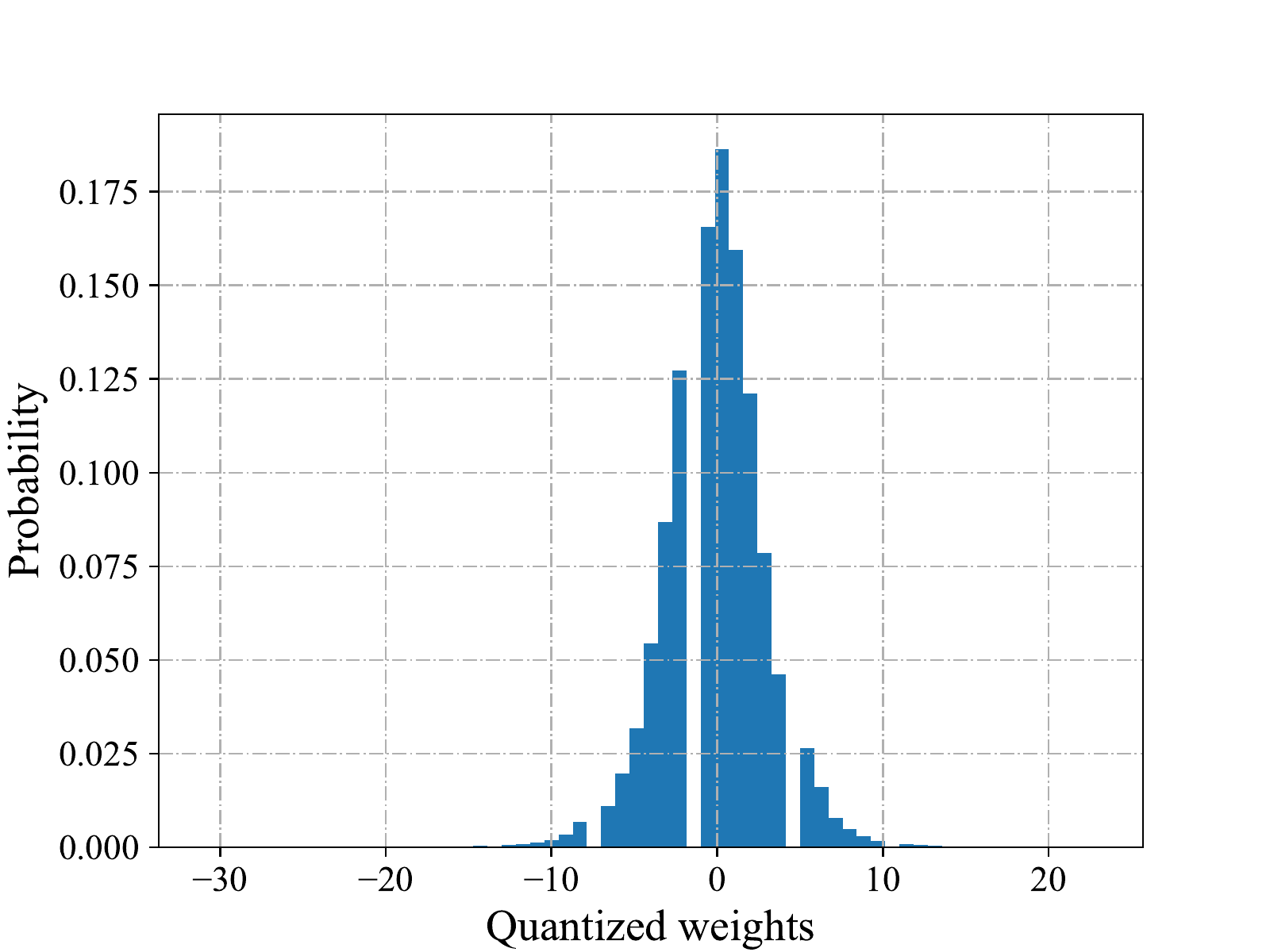}}
\caption{The distribution of quantized weights at 1000 iterations.}
\label{fig:quantized_weights_1000iteration}
\end{figure}

Regarding the redundancy removal within a model, we apply quantization to represent the neural network model in a quantized form. This greatly facilitates efficient compression on these discrete values with ensured performance, since the quantization operation has already been embedded in the network during the training process. 
In our method, the parameter number of \begin{math} W_{Q}\end{math} is the same as \begin{math} W_{L}\end{math} and  \begin{math} W_{conv}\end{math}.
We only quantize the reparameterizations of convolutional weights which account for a large proportion of the model representation, as illustrated in Tables~\ref{tab:Parameter_number_comparison},~\ref{tab:paremeter_setting_bias} and~\ref{tab:paremeter_setting_Weights}. 
The number of parameters influences the model size and performance. The model with more filters achieves better recovery performance with the expense of higher representation overhead. As such, a trade-off between number of parameters and the recovery performance is expected.
We adopt the 32-bit floating point format to represent \begin{math} W_{L}\end{math} for accurate learning while the quantized parameters \begin{math} W_{Q}\end{math} are in a 32-bit integer format.

In theory, the number of bits for representing \begin{math} W_{L}\end{math} and  \begin{math} W_{Q}\end{math} is identical, while practically the average number of bits can be reduced by the quantization process. In a common sense, the average number of bits for representing 32-bit floating number is 32. The underlying reason is that the distribution of floating number \begin{math} W_{L}\end{math} during training is continuous, and it is 
quite rare that two floating numbers share the same value. 
However, the distribution of the quantized weights \begin{math} W_{Q}\end{math}  is discrete as shown in  Fig. \ref{fig:quantized_weights_1000iteration}  and the average number of bits for representation depends on the entropy of the discrete signal. 
To quantitatively quantify the reduction of redundancy within a model by  quantization, the entropy of the quantized parameters \begin{math} W_{Q}\end{math} at 30000 iterations is calculated, which is 2.31 while the number of bits to represent \begin{math} W_{L}\end{math} is 32. 
In this manner, it is estimated that the compression ratio of 13.85 times can been achieved, taking the sequence of ``BQTerrace'' as an example.

\begin{figure}[t]
\centering
{\includegraphics[width = 0.35\textwidth]{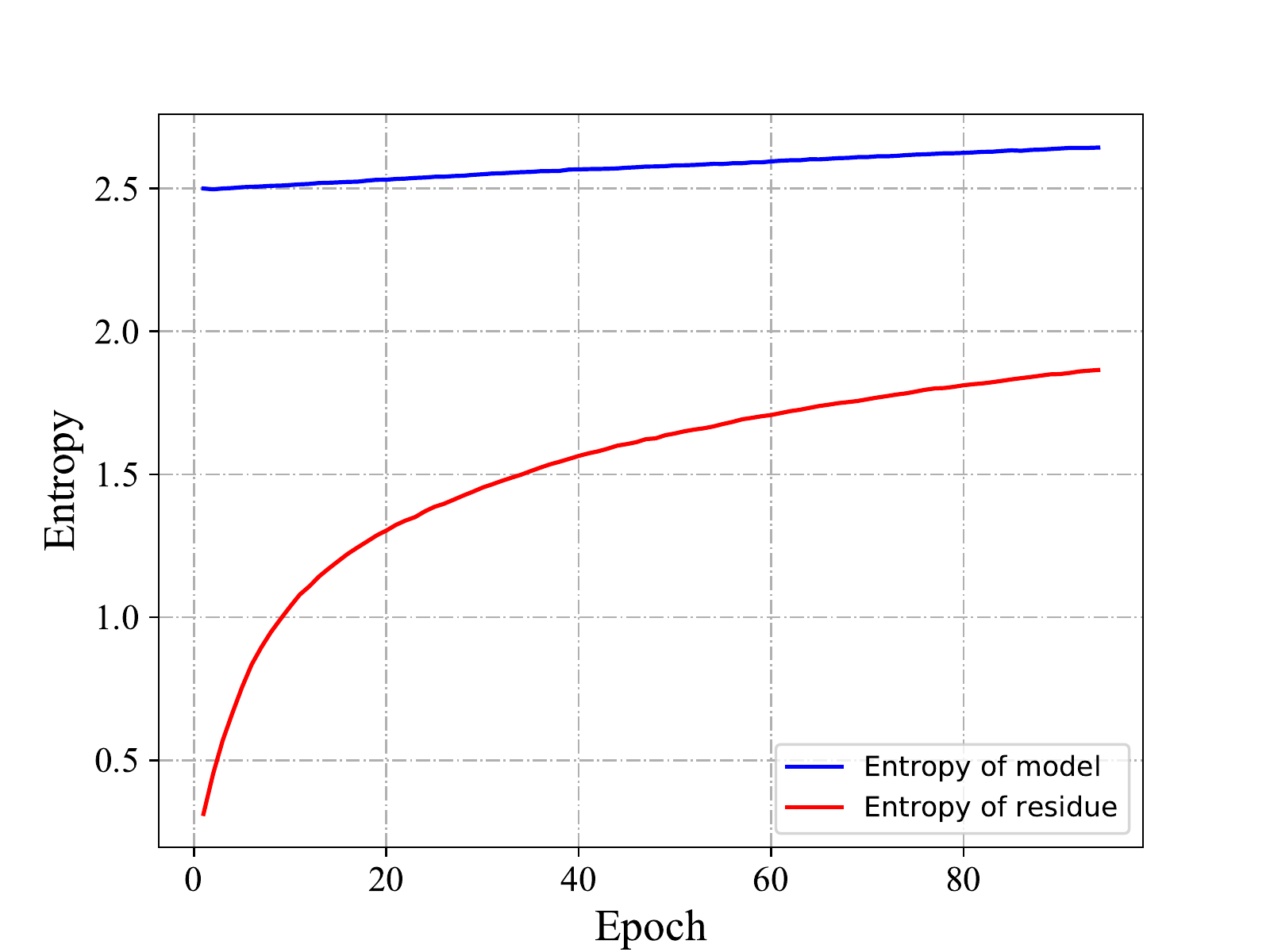}}
\caption{Entropy comparisons between the original model and residue.}
\label{fig:Entropy_comparison}
\end{figure}

Regarding the redundancy existing between different models, the residue between online learned quantized model and initial quantized model is compressed by arithmetic coding, which can further economize the model transmission cost and achieve better compression efficiency. 
The inherent reason lies in the fact that weights of current model are learned by fine-tuning the initial model such that there exist high correlations. 
To illustrate the efficiency of removing the inter model redundancies, the entropy of residue and quantized parameters \begin{math} W_{Q}\end{math} are presented in Fig. \ref{fig:Entropy_comparison}. 
It can be found that the entropy of residue is increasing with the epoch due to larger difference between learned model at each iteration and the initial model. 
It is also interesting to see that the entropy of residue is increasing fast at the beginning of online learning and converges after a few epochs (here we set 1000 iterations as one epoch). Moreover, the entropy of the learned model is much higher than the corresponding residue at each iteration. 
The residue compressed by arithmetic coding is transmitted along with the BL to the receiver side.

To illustrate the model compression performance, the compression ratio is further calculated at each epoch. 
It is worth mentioning that the biases and the weights of affine transform are not quantized, and the raw values in float32 format are incorporated into the final bitstream. Therefore, the total model stream includes the compressed residue of quantized weights,  the biases and the weights of affine transform. More specifically, the fixed size of biases and weights of affine transform is 1.348KB and 0.368KB, respectively.
The compression ratio is formulated as follows,
\begin{equation}
 \label{eq:compression_ratio}
   Ratio = \frac{R_{ori}}{R_{model}}, \\
\end{equation}
where
\begin{equation}
 \label{eq:compression_ratio}
   R_{model} = R_{res} + R_{Biases}+R_{f_{\varphi}}. \\
\end{equation}
Here, \begin{math}R_{ori}\end{math} and \begin{math} R_{model}\end{math} indicate the number of bits of the original uncompressed model and the compressed model, respectively. \begin{math}R_{res}\end{math}, \begin{math}R_{Biases}\end{math} and \begin{math}R_{f_{\varphi}}\end{math} indicate the number of bits of residue, biases and the weights of affine transform, respectively.

\begin{figure}[t]
\centering
{\includegraphics[width = 0.40\textwidth]{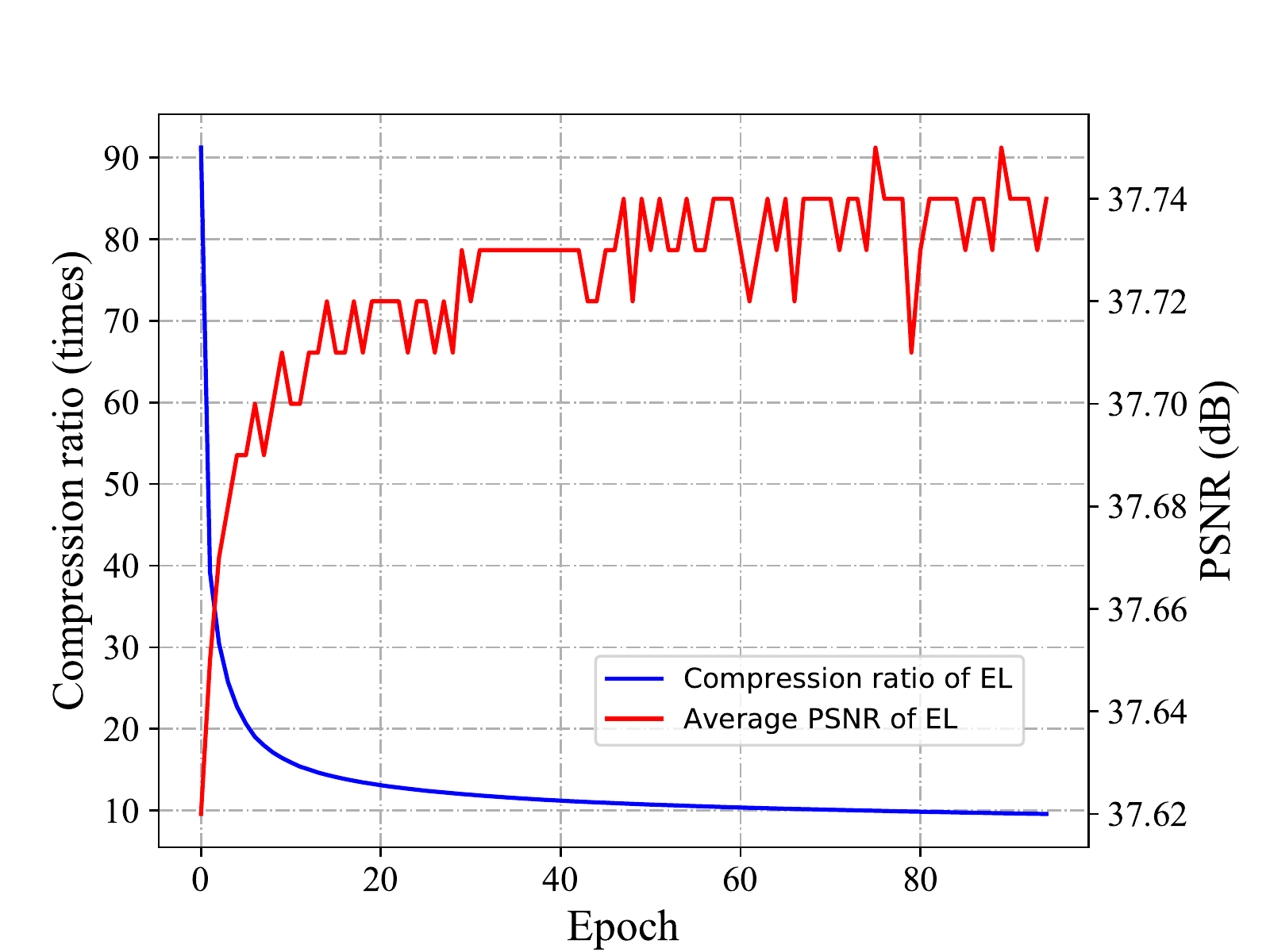}}
\caption{Compression ratio $vs$ PSNR.}
\label{fig:compression_ratio_PSNR}
\end{figure}

As shown in Fig.~\ref{fig:compression_ratio_PSNR},  at the beginning of online learning, we can achieve relatively high compression ratio, due to the fact that the difference between the current model and initial model is relatively small. 
With the decrease of compression ratio, the recovery capability becomes better in the early stage of online learning. 
However, after several epochs, both the compression ratio  and performance of recovery converge. 
Although there are models with promising performance by matching the original signals, the final performance governed by both recovery capability and overhead of EL may not be satisfactory. 
Consequently, the selection of models becomes critical, which further motivates us to design the model selection scheme based on rate-utility optimization.

\subsection{Rate-utility optimization}
Rate-utility optimization aims to figure out an optimally compressed model that achieves the best trade-off between model rate and utility. Herein, the utility is defined based on the final utility of the EL, i.e., recovering the visual signal. As such, instead of the distortion of the compressed model, the quality of visual signals is what matters. 
In our scheme, after model compression, the candidate compressed models are  obtained, which are subjected to be evaluated by  
the Lagrangian cost,
\begin{equation}
 \label{eq:quantizeProcess}
   J = \sum_{i=1}^{N}J_{i},
\end{equation}
and
\begin{equation}
 \label{eq:quantizeProcess}
   J_{i}=D_{i}+\lambda_{i} \times (R_{i}+\frac{R_{model}}{N}),
\end{equation}
where \begin{math} J_{i}\end{math} indicates the Lagrangian cost of $i^{th}$ frame after enhancement and \begin{math} N \end{math} is the number of frames in the group.  \begin{math}D_{i}\end{math} indicates the Sum of the Squared Error (SSE) of $i^{th}$ enhanced  frame by the selected model,\begin{math} ~\lambda_{i}\end{math} indicates the
Lagrange multiplier of $i^{th}$ frame. \begin{math} ~R_{i} \end{math} indicates the number of bits of $i^{th}$ frame encoded  in BL,  and  \begin{math}R_{model}\end{math} is the number of bits of selected model. 
It is worth mentioning that the cost of model rate is assigned to each frame.
Finally, the model with the minimum Lagrangian cost is selected and forms as the EL.

\begin{figure}[t]
\centering
{\includegraphics[width = 0.40\textwidth]{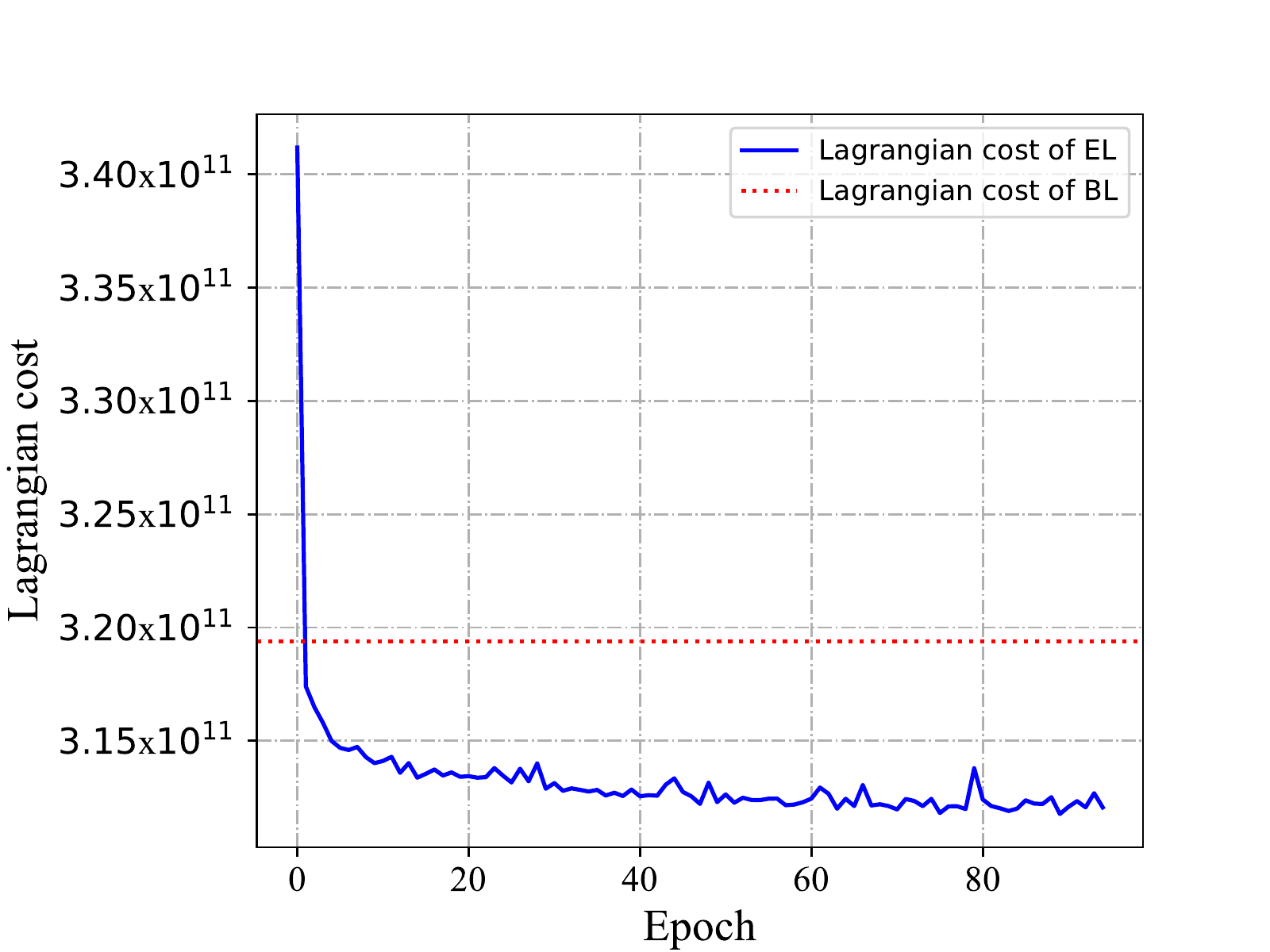}}
\caption{Lagrangian cost during online learning.}
\label{fig:Lcost_qq22_BQtre}
\end{figure}

Taking the sequence of ``BQTerrace'' as an example,  the Lagrangian cost of model at each epoch is shown in Fig. \ref{fig:Lcost_qq22_BQtre}. 
The red line represents the Lagrangian cost of the BL and the first point of the curve indicates the Lagrangian cost of the initial model. 
It can be found that in this case, the Lagrangian cost decreases with the increasing of the epoch, finally leading to better representation performance. It is also interesting to see that the cost of the initial model is worse than the BL due to the fact that the recovery performance may not always be satisfactory when the adaptation to the specific content is absence.

\begin{table}[t]
  \caption{Performance comparisons in terms of BD-Rate under AI configuration (anchor: VTM-4.0).}
  \label{tab:luma_performance_comparison_AI}
  \begin{tabular}{cccc}
  \toprule
    Sequences & JVET-N0254  & JVET-N0513 & Proposed\\
     \midrule
Class A &   -1.22\%	&-0.39\%&	-1.94\% \\
Class B &   -0.93\%	&-0.58\%&	-2.51\% \\
Class C & -1.90\%&	-1.63\%&	-3.65\%\\
Class D &-2.57\%&	-1.32\%&	-2.08\%\\
Class E &-2.22\%&	-2.15\%	&-5.34\%\\
Class F &-1.09\%&	-1.00\%&	-4.90\%\\
 \hline
Overall &-1.56\%&	-1.09\%	&-2.63\%\\

 \bottomrule
\end{tabular}
\end{table}

\section{Experimental results and analyses}

\begin{table*}
\renewcommand{\arraystretch}{1.1}
  \caption{Performance comparisons in terms of BD-Rate under RA configuration (anchor: VTM-4.0).}
  \label{tab:luma_performance_comparison_RA}
  \begin{tabular}{cccccc}
    \toprule
    Sequences & JVET-N0110 & JVET-N0254 & JVET-N0480 & JVET-N0513 & Proposed\\
     \midrule
    Class A   & -2.21\%  & -1.74\%  & -1.06\%  &-0.37\%  & -3.21\% \\
    Class B   & -1.52\%  & -1.13\%  & -0.55\%  &-0.43\%  & -4.64\% \\
    Class C   & 0.12\%   & -1.39\%  &  0.09\%  &-0.76\%  &-4.60\%   \\
    Class D   &  -        & -1.39\%  & -         &-0.79\%  & -4.50\%   \\
    Class F   & -         & -0.50\%  & -          &-0.35\%  & -3.70\%   \\
    \hline
    Overall   & -1.36\%  & -1.27\%  & -0.58\%  &-0.52\%  & -4.07\%   \\
 \bottomrule
\end{tabular}
\end{table*}

\subsection{Experimental setup}
To evaluate the performance of our method, the reference software of the upcoming video coding standard VVC (VVC Test Model version 4.0, VTM-4.0)   is incorporated as the BL.
The initial model is trained by using the database of DIV2K \cite{Agustsson_2017_CVPR_Workshops}, which consists of 900 PNG pictures with the resolution of 2K (800 images for training and 100 images for validation). 
To facilitate the comparison with other methods, the video sequences are compressed  with All Intra (AI) and Random Access (RA) configurations under Common Test Conditions (CTC) \cite{VVC-CTC}.  
The learned models are applied on the luminance channel and the  Quantization Parameters (QPs) are set following CTC \{22, 27, 32, 37\}. 

The module of online learning in EL is implemented with tensorflow software package~\cite{tensorflow2015-whitepaper}. All the frames of input video and corresponding reconstructed frames are cropped into patches randomly and the patch size is set as $35 \times 35$. The learning rate is set as 0.0002.  
It should be noted that the online learning will terminate if the size is larger than 13.5KB. The scale factor in scale transform is set as 10. 

\begin{figure*}[t]

	\centering
   \subfigure[Original]{\includegraphics[width = 0.33\textwidth]{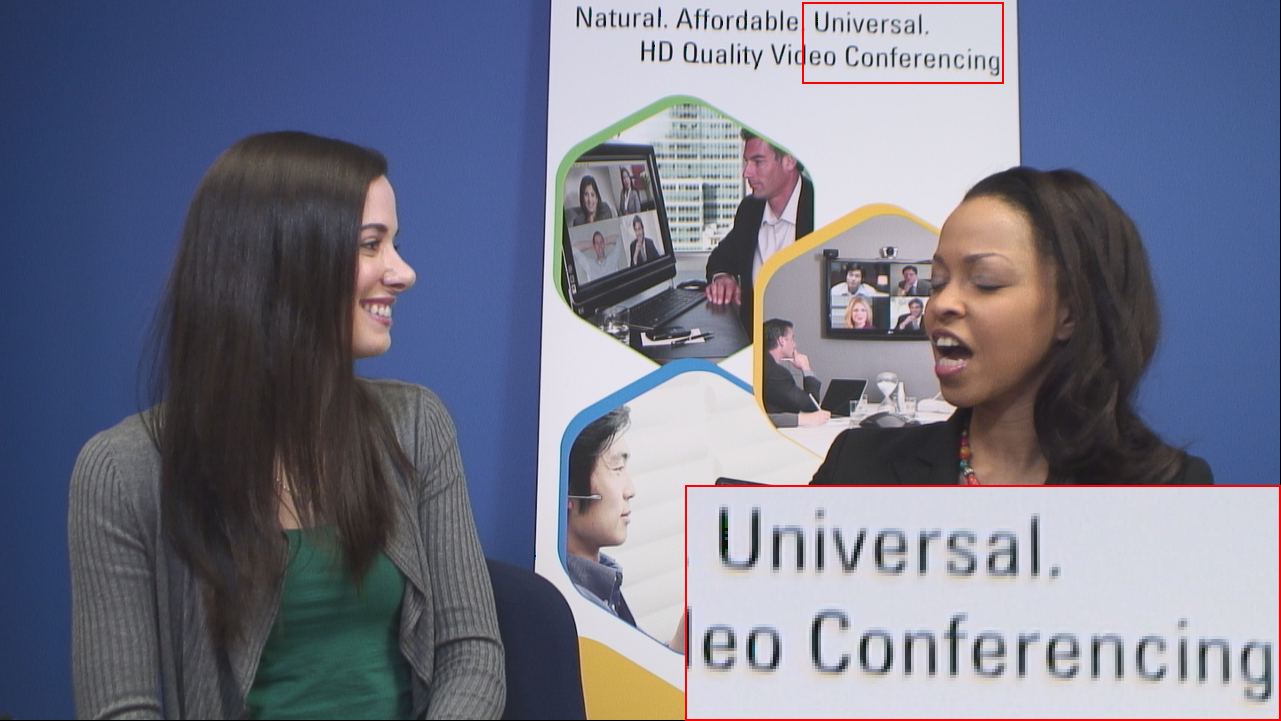}}
   \subfigure[VVC (37.65 dB) ]{\includegraphics[width = 0.33\textwidth]{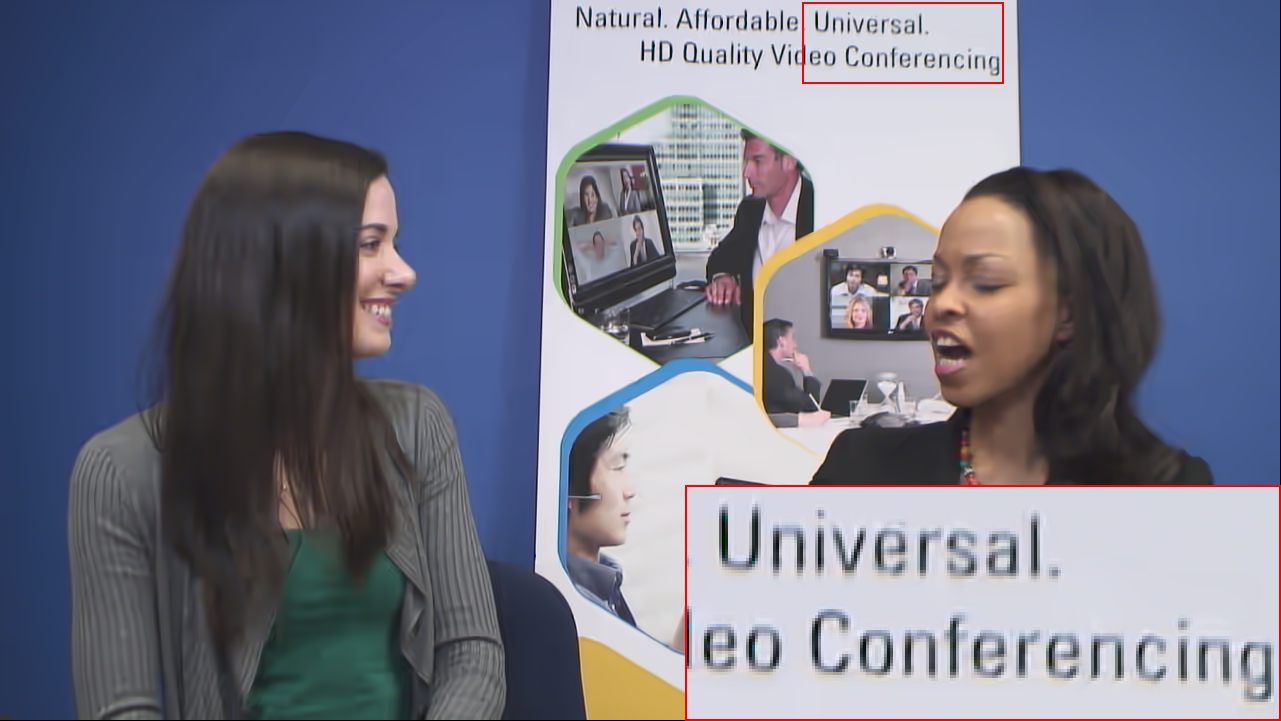}}
   \subfigure[Proposed (38.18 dB)]{\includegraphics[width = 0.33\textwidth]{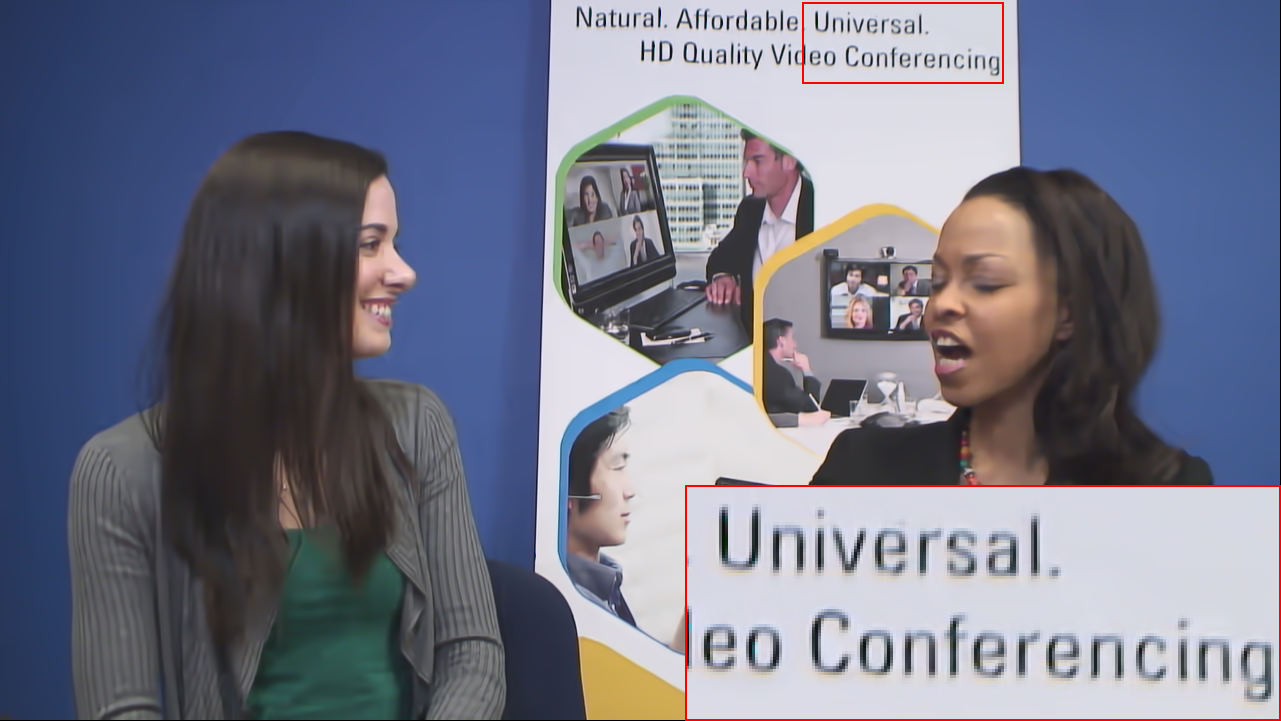}}
   
    \caption{Visual quality comparisons for ``KristenAndSara'' under AI configuration, where the 241-th frame is shown (QP=37).}
    \label{fig:Subject_evalution_AI}
\end{figure*}

\begin{figure*}[t]

	\centering
   \subfigure[Original]{\includegraphics[width = 0.33\textwidth]{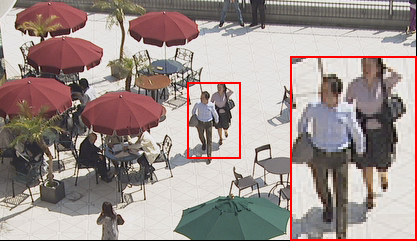}}
   \subfigure[VVC (30.40 dB) ]{\includegraphics[width = 0.33\textwidth]{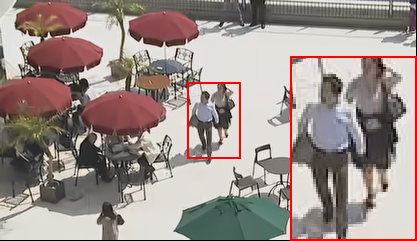}}
   \subfigure[Proposed (31.32 dB)]{\includegraphics[width = 0.33\textwidth]{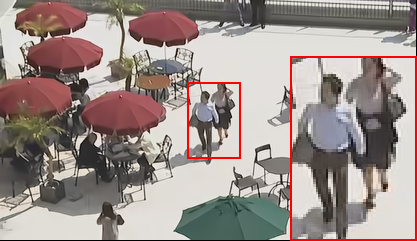}}
   
    \caption{Visual quality comparisons for ``BQSquare'' under RA configuration, where the 184-th frame is shown (QP=37).}
     \label{fig:Subject_evalution_RA}
\end{figure*}

\subsection{Performance comparisons}
In this section, the proposed method is
compared with the state-of-the-art algorithms that enhance the quality of decoded videos on the platform of VVC, including JVET-N0110~\cite{JVET-N0110}, JVET-N0254~\cite{JVET-N0254},  JVET-N0480~\cite{JVET-N0480}  and JVET-N0513~\cite{JVET-N0513}. The coding performance is measured by BD-Rate~\cite{bjontegaard2001calculation} and the anchor is VTM 4.0. 
In particular, in JVET-N0254 and JVET-N0513, a dense residual CNN and two light weight deep CNNs are learned by the offline learning scheme. 
The performance of the offline learning based methods highly depends on the training dataset while our proposed method can well adapt to the variation on video content.
In JVET-N0110 and JVET-N0480, online learning scheme is utilized to achieve an adaptive CNN loop filter and the parameters of the learned model are signaled to the decoder side. 
Compared to their methods, in our scheme the model compression is naturally incorporated in the learning process, and the final representation of the model is selected in a rate-utility optimization sense, leading to the improvement of the performance.

From  Table \ref{tab:luma_performance_comparison_AI}, it is observed that the proposed method achieves 2.63\% bit-rate savings on average under AI configuration while the methods of JVET-N0254 and JVET-N0513 achieve 1.56\% and  1.09\% bit rate reductions, respectively.  From  Table \ref{tab:luma_performance_comparison_RA}, our proposed method achieves 4.07\%  bit-rate savings on average under RA configuration while the methods of JVET-N0110, JVET-N00254,  JVET-N0480, and JVET-N0513 reduce 1.36\%,  1.27\%, 0.58\% and  0.52\% bit rate on average, respectively. It can be easily found that our proposed method outperforms these CNN based algorithms under both AI and RA configurations. In Table \ref{tab:luma_performance_comparison_AI}, it is also observed that the performance of class D of our proposed is marginally worse than the method of JVET-N0254.
The reason is that the resolution of sequences in class D is $416\times240$, which is smaller than that of other sequences. As such, the relatively smaller bit rates of video streams in BL lead to higher percentage of the overhead bit rate in EL.  
Moreover, under AI configuration only one frame in a group (8 frames) is encoded such that the relative overhead bit rate (model stream) is much higher than RA configuration even the size of model is very close. 
However, as shown in Table \ref{tab:luma_performance_comparison_RA}, our method outperforms other methods in all classes under RA configuration because the overhead bit rate in EL is assigned to all frames in a group. To further demonstrate the relationship between PSNR gain and the bit rate in EL, the results of each sequence in class B are presented in Table \ref{tab:luma_performance_proposed_RA}. 
It can be found that for the same sequence, the model size is different under different QP settings since our method selects the optimal model with minimum rate-utility cost.

\begin{table*}
\renewcommand{\arraystretch}{1.06}
  \caption{The coding performance and corresponding model bitrate for each sequence of CLASS B under RA configuration (anchor: VTM-4.0).}
  \label{tab:luma_performance_proposed_RA}
  \begin{tabular}{cccccccc}
    \toprule
    Sequences  & Frame rate &QP &\tabincell{c}{ Bitrate (VTM-4.0)\\ (Kb/s)} &\tabincell{c}{Model size\\ (KB)}  &\tabincell{c}{ Model bitrate\\ (Kb/s)} &  \tabincell{c}{$\Delta$ PSNR\\ (dB)} & BD-Rate  \\
    \midrule
    \multirow{4}*{MarketPlace}   &\multirow{4}*{60} &22  &14252.95&10.95	&8.76	&0.04	&\multirow{4}*{-1.73\%}	\\
		&&27&5426.06&12.06	&9.65	&0.05	\\	
		&&32&2430.38&10.77	&8.61	&0.06	\\
		&&37&1079.80&12.22	&9.77	&0.07	\\	
    \multirow{4}*{RitualDance}   &\multirow{4}*{60} &22 &9443.66&11.16	&8.93	&0.10
	&\multirow{4}*{-2.20\%}	\\
		&&27&4717.00&11.39	&9.11	&0.12\\	
		&&32&2514.74&11.01	&8.80	&0.12\\
		&&37&1322.51&11.20	&8.96	&0.12	\\	
    \multirow{4}*{Cactus} &\multirow{4}*{50}  &22 &14402.89&12.49	&9.99	&0.05 &\multirow{4}*{-4.48\%}	\\
	&&27&4300.21&12.13&	9.71&	0.09\\
    &&32&1998.80&12.56	&10.06	&0.12\\
    &&37&971.69&12.58	&10.06	&0.16\\
     \multirow{4}*{BasketballDrive}   &\multirow{4}*{50} &22 &14684.65&12.05	&9.64	&0.04
     &\multirow{4}*{-4.33\%}	\\
	&&27&4788.42&12.25&	9.80&	0.09\\
    &&32&2235.44&12.23&	9.79&	0.13\\
    &&37&1117.68&12.48&	9.98&	0.15\\
    \multirow{4}*{BQTerrace}   &\multirow{4}*{60} &22 &34653.52 &13.06	&10.45	&0.10 &\multirow{4}*{-10.45\%}	\\
&&27&5827.09&12.73&	10.18&	0.15\\
&&32&1765.12&12.37&	9.89&	0.18\\
&&37&779.22&12.27&	9.82&	0.22\\

   \bottomrule
\end{tabular}
\end{table*}

Regarding the subjective quality comparisons, the original frames, VVC decoded frames, and reconstructed frames from the proposed method for ``KristenAndSara'' and ``BQSquare'' sequences are shown in Figs. \ref{fig:Subject_evalution_AI} and \ref{fig:Subject_evalution_RA}. 
For better visualization, certain regions are also enlarged. 
It can be observed that the ringing artifacts and blocking artifacts are eliminated when compared with the anchor. 
The degraded structural details have also been well recovered since our method can well leverage the deep neural network representation to accommodate the statistics of the visual signals. More specifically, when enlarging Figs. \ref{fig:Subject_evalution_AI} (b) and (c), it can be observed the marginal pixels of “U” in (b) are mixed with unexpected white pixels of background. In (c), these white pixels are suppressed. In Figs. \ref{fig:Subject_evalution_RA} (b) and (c), scrupulous observers may also find that the area around the woman’s left hand in (c) is smoother than the area in (b). 
Due to powerful transferable capability with acceptable rate overhead, the proposed method achieves effectively improved visual information representation performance.

\section{Conclusions}

In this paper, a novel scheme for visual signal representation that leverages  transferable modality has been proposed. 
In particular, 
online learning that accommodates the statistics of input signals serves as a transferable engine from visual information to neural network model which is further compactly represented via inter/intra model reduduancy removal. 
The trade-off between rate and utility is further optimized, leading to the best representation capability.  
With the state-of-the-art video coding platform of VVC, extensive experiments show that  visual information capability has been improved with significant bit rate savings.

\begin{acks}
This work was supported in part by the Hong Kong RGC General Research Funds under Grant 9042322 (CityU 11200116), Grant 9042489 (CityU 11206317), and Grant 9042816 (CityU 11209819).
This work was also supported in part by the Natural Science Foundation of China under Grant 61672443, Grant 61901459, and in part by China Postdoctoral Science Foundation under Grant 2019M653127.
\end{acks}


\bibliographystyle{ACM-Reference-Format}
\balance
\bibliography{sample-base}

\end{document}